\documentclass[prl,reprint,aps,amsmath,amssymb,twocolumn,superscriptaddress,longbibliography]{revtex4-1}

\usepackage{color}
\usepackage{graphicx}% Include figure files
\usepackage{dcolumn}% Align table columns on decimal point
\usepackage{bm}% bold math
\usepackage{hyperref,xcolor,times}% add hypertext capabilities
\usepackage{siunitx}
\usepackage{diagbox}

\definecolor{linkblue}{rgb}{0, 0, 1}
\hypersetup{
	colorlinks=true,       		% false: boxed links; true: colored links
	linkcolor=linkblue,          	% color of internal links
	citecolor=linkblue,             % color of links to bibliography
	urlcolor=linkblue,          % color of external links
}

\begin{document}

\title{Purcell-Enhanced Generation of Photonic Bell States via the Inelastic Scattering of Single Atoms}
	
\author{Jian Wang}
\email{jwang28@ustc.edu.cn}
\affiliation{CAS Key Laboratory of Quantum Information, University of Science and Technology of China, Hefei 230026, China}
\affiliation{Anhui Province Key Laboratory of Quantum Network, University of Science and Technology of China, Hefei 230026, China}
\affiliation{CAS Center For Excellence in Quantum Information and Quantum Physics, University of Science and Technology of China, Hefei 230026, China}

\author{Xiao-Long Zhou} 
\thanks{These authors contributed equally to this work.}
\affiliation{CAS Key Laboratory of Quantum Information, University of Science and Technology of China, Hefei 230026, China}
\affiliation{Anhui Province Key Laboratory of Quantum Network, University of Science and Technology of China, Hefei 230026, China}
\affiliation{CAS Center For Excellence in Quantum Information and Quantum Physics, University of Science and Technology of China, Hefei 230026, China}

\author{Ze-Min Shen}
\affiliation{CAS Key Laboratory of Quantum Information, University of Science and Technology of China, Hefei 230026, China}
\affiliation{Anhui Province Key Laboratory of Quantum Network, University of Science and Technology of China, Hefei 230026, China}
\affiliation{CAS Center For Excellence in Quantum Information and Quantum Physics, University of Science and Technology of China, Hefei 230026, China}

\author{Dong-Yu Huang}
\affiliation{CAS Key Laboratory of Quantum Information, University of Science and Technology of China, Hefei 230026, China}
\affiliation{Anhui Province Key Laboratory of Quantum Network, University of Science and Technology of China, Hefei 230026, China}
\affiliation{CAS Center For Excellence in Quantum Information and Quantum Physics, University of Science and Technology of China, Hefei 230026, China}
\affiliation{Hefei National Laboratory, University of Science and Technology of China, Hefei 230088, China}

\author{Si-Jian He}
\affiliation{CAS Key Laboratory of Quantum Information, University of Science and Technology of China, Hefei 230026, China}
\affiliation{Anhui Province Key Laboratory of Quantum Network, University of Science and Technology of China, Hefei 230026, China}
\affiliation{CAS Center For Excellence in Quantum Information and Quantum Physics, University of Science and Technology of China, Hefei 230026, China}

\author{Qi-Yang Huang}
\affiliation{CAS Key Laboratory of Quantum Information, University of Science and Technology of China, Hefei 230026, China}
\affiliation{Anhui Province Key Laboratory of Quantum Network, University of Science and Technology of China, Hefei 230026, China}
\affiliation{CAS Center For Excellence in Quantum Information and Quantum Physics, University of Science and Technology of China, Hefei 230026, China}

\author{Yi-Jia Liu}
\affiliation{CAS Key Laboratory of Quantum Information, University of Science and Technology of China, Hefei 230026, China}
\affiliation{Anhui Province Key Laboratory of Quantum Network, University of Science and Technology of China, Hefei 230026, China}
\affiliation{CAS Center For Excellence in Quantum Information and Quantum Physics, University of Science and Technology of China, Hefei 230026, China}
\affiliation{Hefei National Laboratory, University of Science and Technology of China, Hefei 230088, China}

\author{Chuan-Feng~Li}
\email{cfli@ustc.edu.cn}
\affiliation{CAS Key Laboratory of Quantum Information, University of Science and Technology of China, Hefei 230026, China}
\affiliation{Anhui Province Key Laboratory of Quantum Network, University of Science and Technology of China, Hefei 230026, China}
\affiliation{CAS Center For Excellence in Quantum Information and Quantum Physics, University of Science and Technology of China, Hefei 230026, China}
\affiliation{Hefei National Laboratory, University of Science and Technology of China, Hefei 230088, China}

\author{Guang-Can~Guo}
\affiliation{CAS Key Laboratory of Quantum Information, University of Science and Technology of China, Hefei 230026, China}
\affiliation{Anhui Province Key Laboratory of Quantum Network, University of Science and Technology of China, Hefei 230026, China}
\affiliation{CAS Center For Excellence in Quantum Information and Quantum Physics, University of Science and Technology of China, Hefei 230026, China}
\affiliation{Hefei National Laboratory, University of Science and Technology of China, Hefei 230088, China}

%\date{\today}

\begin{abstract}
Single atoms trapped in optical cavities exhibit immense potential as key nodes in future quantum information processing. They have already demonstrated significant advancement in various quantum technologies, particularly regarding the generation of nonclassical light. Here, we efficiently produce genuine photonic Bell states through the inelastic scattering process of single two-level intracavity atoms. An experimental violation of the Bell inequality, arising from the interference between the probability amplitudes of two photons, validates the intrinsic nature of energy-time entanglement. Coupling atoms with an optical cavity in the Purcell regime substantially enhances the two-photon scattering. This Bell state generation process does not require atomic spin control, thereby rendering it inherently immune to decoherence effects. This work advances the comprehension of resonance fluorescence and has the potential to broaden the landscape of quantum technologies and facilitate the application of photonic Bell states.
\end{abstract}

\maketitle

Nonlinearity in quantum systems reveals the complexities of quantum interactions and is eagerly sought after in quantum technologies \cite{chang2014quantum, peyronel2012quantum}. Atomic systems have demonstrated significant experimental advancements, including photon blockades \cite{birnbaum2005photon, hamsen2017two}, photon bound states \cite{liang2018observation}, single-photon switches \cite{shomroni2014all}, and photon-photon quantum gates \cite{hacker2016photon}. The nonlinear response of an atomic system is governed by the strength of its interaction with the surrounding optical field. Nonetheless, the spatial extent of the optical field is typically several orders of magnitude greater than the atomic absorption cross-section, leading to a substantially weak atom-photon coupling. By coupling a single atom with a high-finesse optical cavity, the atom-photon interaction can be significantly enhanced through repeated photon reflections. This enhancement makes a single intracavity atom an ideal platform for conducting a quantum nonlinearity investigation \cite{covey2023quantum}. When the \textit{cooperativity} $C$ described by the cavity quantum electrodynamics (QED) of an atom-cavity system is sufficiently high, the single atom will interact deterministically with photons with a probability approaching unity \cite{reiserer2015cavity}. Experimentally, by aligning the resonance of the optical cavity with the relevant transition processes, a system can be engineered to evolve according to the targeted dynamical process \cite{reiserer2022colloquium}.

Resonance fluorescence, as one of the pioneering and still vigorously studied topics in quantum optics, has inspired both theoretical and experimental investigations into nonclassical light \cite{kimble1976theory, wu1975investigation,  kimble1977photon, aspect1980time, ng2022observation, liu2024dynamic,  lopez2024entanglement}. Nonclassical features such as antibunching and squeezing arise from the nonlinearity caused by the inelastic scattering of single atoms with resonant light \cite{walls1981reduced, del2012theory, hanschke2020origin}. The Mollow triplet is a spectral signature of the scattered light field derived from a two-level atom subjected to resonant optical driving \cite{mollow1969power, ortiz2019mollow}. The antibunching of resonance fluorescence has been demonstrated to originate from the quantum interference between the elastic and inelastic components \cite{phillips2020photon, casalengua2024photons}. A more profound investigation have shown that in the Heitler regime, the resonance fluorescence of two-level atoms features simultaneous scattering of two photons\cite{masters2023simultaneous}. Recently, a pioneering  and innovative study based on quantum dots has rigorously demonstrated the entanglement phenomenon inherent in resonance fluorescence by violation of Bell inequality \cite{liu2024violation}.  As a fundamental and significant discovery, this entanglement phenomenon warrants exploration in additional physical systems. Natural atoms, suffering from narrow linewidths, low photon collection efficiency, and difficulties in trapping, have not yet observed this phenomenon \cite{ng2022observation, masters2023simultaneous}.

%%%%%%%%%%%%%%%%%%%%%%%%%%%%%%%%%%%%%%%%%%%%%%%%%%%%%%%%%%%%%%%%%%%
\begin{figure*}[tb]
    \centering
        \includegraphics[width=1.0\textwidth]{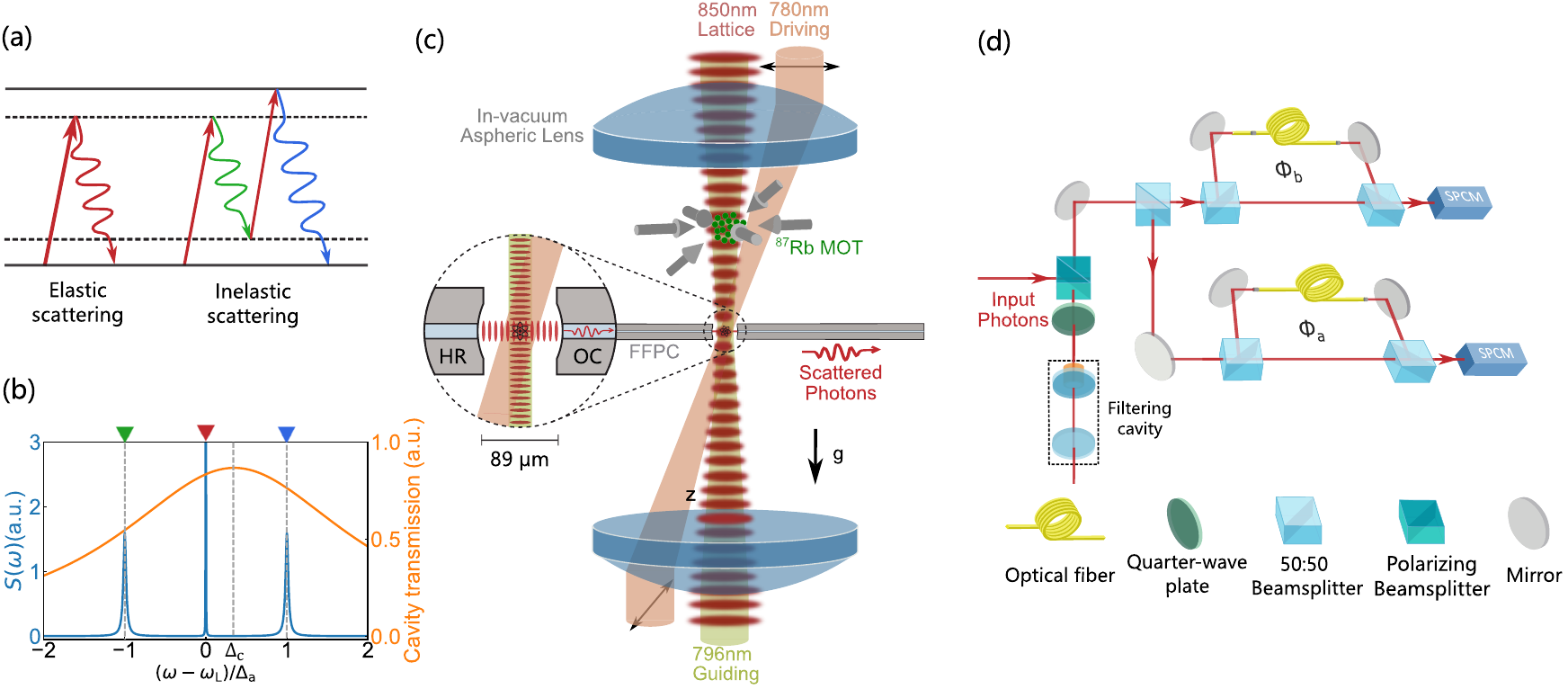}
\caption{
  Principle diagram of two-photon inelastic scattering process and experimental setup. (a) Elastic and inelastic scattering processes of a two-level single atom in Heitler regime. (b) Spectrum of resonance fluorescence in Heitler regime and cavity transmission. The ratio of inelastic to elastic scattering depends on the saturation parameter. (c) A single atom is loaded into the FFPC which is composed of a high-reflective (HR) and an out-coupling (OC) mirror, and is trapped in a two-dimensional optical lattice. The driving light is focused into the cavity region via two in-vacuum lenses. (d) The inelastically scattered photons are frequency-filtered out by a narrowband notch filter and Bell inequality is tested by a Franson interferometer. SPCM, single-photon counting module.}
\label{fig:setup} 
\end{figure*}

%%%%%%%%%%%%%%%%%%%%%%%%%%%%%%%%%%%%%%%%%%%%%%%%%%%%%%%%%%%%%%%%%%%

In this Letter, inspired by the work \cite{liu2024violation}, we unequivocally verify the entanglement phenomenon arising from inelastic scattering in Heitler regime based on a neutral atom-optical cavity system. The coupling between single atoms and optical cavities lies within the Purcell regime \cite{reiserer2015cavity}, which allows for sufficient cooperativity while simultaneously enhancing entangled photons at two frequencies. This is nontrivial in natural atomic cavity QED systems. Experimentally, a fiber Fabry-Pérot cavity (FFPC) \cite{hunger2010fiber, brekenfeld2020quantum, pfeifer2022achievements, li2023experimental} is employed, which couples single atoms in the Purcell regime, resulting in enhanced photon pair scattering and improved photon collection efficiency. Leveraging this enhancement, the genuine energy-time entanglement is verified through the violation of the Clauser-Horn-Shimony-Holt (CHSH) Bell inequality \cite{clauser1969proposed}, which arises from the correlation between probability amplitudes of two different sideband photons. By deterministically separating two sideband photons with different frequencies, the wavepackets of the entangled photons are measured. Furthermore, continuous tunability of the linewidth of the entangled photons is achieved by detuning the FFPC. This work introduces an approach built on the cavity QED platform to efficiently produce a narrowband photonic Bell state via two-photon inelastic scattering, which does not rely on atomic spin control and is thus immune to decoherence.

We consider a single isolated two-level atom that is exposed to a monochromatic light ($\omega_\mathrm{L}$). Since neutral atoms are nearly impossible to trap for extended periods under strict resonant excitation, the case of large detuning and weak excitation ($\left|\Delta_\mathrm{a}\right|\gg\Omega\gg\Gamma$, Heitler regime) is considered in this Letter, where $\Delta_\mathrm{a}$ is the detuning of the atomic resonance relative to the driving field, $\Omega$ is the Rabi frequency of the driving field and $\Gamma$ is the atomic spontaneous decay rate. At steady state, the atomic population is predominantly in the ground state, with the scattering field comprising both elastic (coherent) and inelastic (incoherent) scattering components. The elastic scattering component inherits the properties of the incident light, exhibiting a delta-function like frequency spectrum in $\omega_\mathrm{L}$, whereas the inelastic scattering component is distributed over a range of frequencies centered at $\omega_\mathrm{L}\pm\Delta_\mathrm{a}$. According to the optical Bloch equation, the ratio of the inelastic component is $R_{\mathrm{sc}}^{(\mathrm{inc})}/ R_{\mathrm{sc}}=s/(s+1)$, which depends on the \textit{saturation parameter} \cite{steck2007quantum}
\begin{equation}\label{equ:sat}
    s=\frac{{2\Omega}^2/\Gamma^2}{1+(2\Delta_\mathrm{a}/\Gamma)^2}.
\end{equation}
The inelastic scattering process can be interpreted as atoms being excited by two-photon terms of the incident field and undergoing cascade radiation mediated by two virtual energy levels \cite{dalibard1983correlation} (Fig.\,\ref{fig:setup}(a)), resulting in an energy-time entangled photon pairs with a frequency of $\omega_\mathrm{L}\pm\Delta_\mathrm{a}$. The total energy of the two photons is conserved, and the emission of one photon must be accompanied by that of the other within a period of $1/\Gamma$, thus they are energy-time entangled \cite{PhysRevA.76.062709, PhysRevLett.121.143601}. 

We further consider a single atom (with a natural linewidth of $2\gamma$) coupled with an optical cavity (with a field decay rate of $\kappa$) with a coupling strength of $g$. In the strong coupling regime ($g\gg\kappa, \gamma$) typically studied in atomic cavity QED experiments, the cavity linewidth is too narrow to simultaneously encompass both sideband photons at two frequencies, leading to a failure in enhancing the two-photon scattering. To address this issue, this work employs an FFPC with a small mode volume, which operates effectively in the Purcell regime ($\kappa>\left(g^2/\kappa\right)>\gamma$). In this regime, the rate at which photons are coherently exchanged between atoms and cavities is much lower than the rate they leak out from cavities, leaving no chance for reabsorption. Moreover, the radiation of both sideband photons can be simultaneously enhanced. In this strong dissipative regime, the coupled system remains degenerate, allowing the atom-cavity interaction to be considered perturbative. As a result, the spontaneous emission rate in Eq.\,\ref{equ:sat} is Purcell broadened to $\Gamma=\left(2C+1\right)\cdot2\gamma$, and the collecting rate of inelastically scattered photons from the optical cavity is $s\cdot R_\mathrm{c}$, where
\begin{equation}\label{equ:R_c}
    R_\mathrm{c}=\frac{2\kappa\Omega^2}{g^2}\frac{\lvert\tilde{C}\rvert^2}{\lvert1+2\tilde{C}\rvert^2},
\end{equation}
is the overall photon collection rate through the cavity emission \cite{gallego2018strong}, which can benefit from higher cavity field decay rate $\kappa$. $\tilde{C}=g^2/\left[2\left(\kappa+i\Delta_\mathrm{c}\right)\left(\gamma+i\Delta_\mathrm{a}\right)\right]$ is the complex cooperativity, $\Delta_\mathrm{c}$ is the detuning of the cavity relative to the driving field \cite{murr2003suppression}.

In our system, as depicted in Fig.\,\ref{fig:setup}(c), a single $^{87}\text{Rb}$ atom ($\gamma/2\pi=\SI{3.0}{\mega\hertz}$) is loaded into an FFPC with a short cavity length of \SI{89.0\pm0.5}{\micro\metre}. A two-dimensional red-detuned optical lattice, formed by a cavity length-stabilizing field of \SI{801}{\nano\metre} and an counter-propgating optical-dipole trap (ODT) of \SI{850}{\nano\metre} tightly focused by two in-vacuum aspheric lenses, confines the atom at an antinode in the center of the cavity mode. These two ODTs induce an AC Stark shift of $\Delta_\mathrm{AC}/2\pi=\SI{13.7\pm0.9}{\mega\hertz}$ on the $F=2\leftrightarrow F'=3$ transition of the $\mathrm{D}_2$ line at \SI{780}{\nano\metre} that we employed. The utilized FFPC has a high field decay rate of $\kappa/2\pi=\SI{164\pm5}{\mega\hertz}$, which is capable of enhancing entangled photons at two frequencies (Fig.\,\ref{fig:setup}(b)) and allowing them to leak out from the cavity at a faster rate.Nevertheless, the small mode volume of the FFPC enables a high measured atom-cavity coupling strength of $g/2\pi=\SI{63\pm5}{\mega\hertz}$, still locating our system well in the Purcell regime with a cooperativity of $C=g^2/2\kappa\gamma=\SI{4.1\pm0.8}{}$. Unlike the work that couples photons into the system via the waveguide \cite{liu2024violation}, this work injects light from the side of FFPC to directly drive the single atoms. The driving light, with a small angle perpendicular to the cavity axis, is red-detuned to $\lvert\Delta_\mathrm{a}\rvert/2\pi=\SI{93.7\pm0.9}{\mega\hertz}$ with respect to the Stark-shifted atomic resonance, with a Rabi frequency of $\Omega/2\pi=\SI{32.2\pm0.8}{\mega\hertz}$ corresponding to a saturation parameter of $s=\SI{0.054\pm0.004}{}$. It is composed of two counter-propagating beams with orthogonal polarizations for balancing the optical pressure exerted on the atom.

%%%%%%%%%%%%%%%%%%%%%%%%%%%%
\begin{figure}[tb]
    \centering 
        \includegraphics[width=0.48\textwidth]{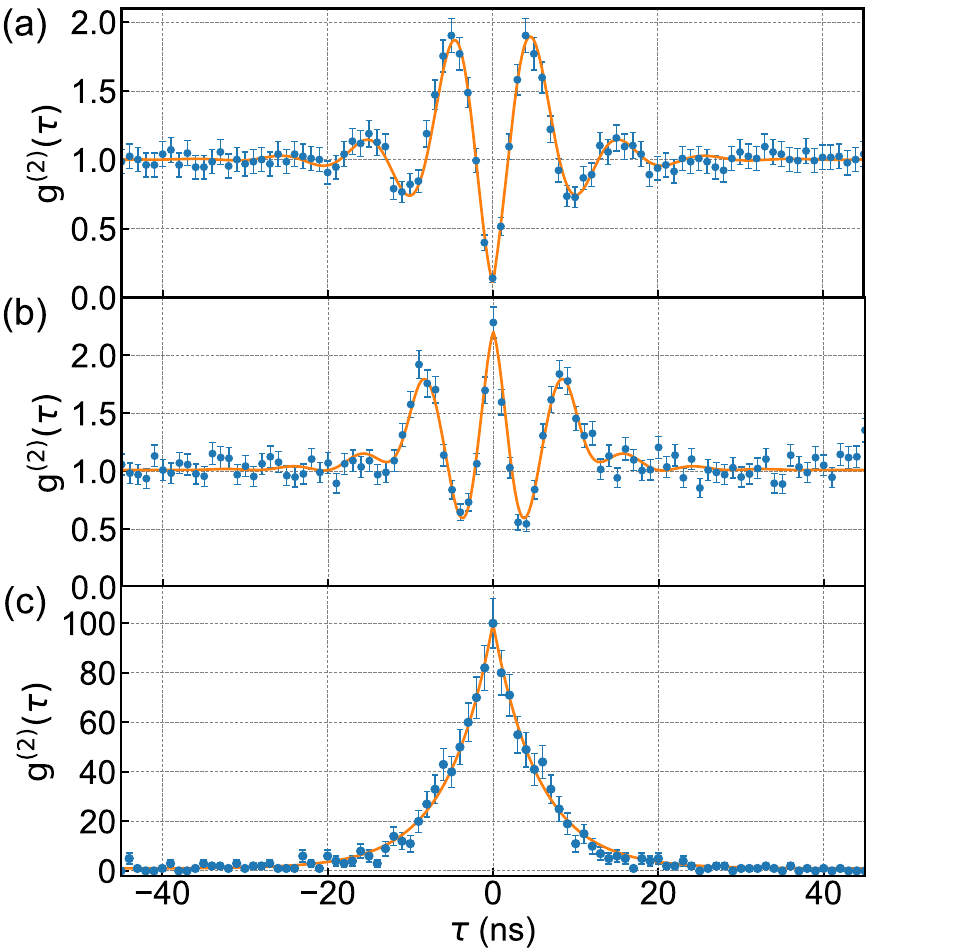}
\caption{Second-order correlation function of the photons collected through the cavity emission with $\Delta_\mathrm{c}/2\pi=\SI{330}{\mega\hertz}$. The measured field contains (a) inelastic and complete elastic scattering photons, (b) inelastic and 50\% filtered elastic scattering photons} and (c) only inelastic scattering photons.

\label{fig:HBT} 
\end{figure}
%%%%%%%%%%%%%%%%%%%%%%%%%%%%%

Once an atom is trapped in the FFPC, it is exposed to the red-detuned driving light and can maintain a relatively long lifetime ($\tau_{1/\mathrm{e}}\approx\SI{20}{\second}$) in this Heitler regime, benefiting from the polarization gradient cooling effect \cite{nussmann2005vacuum}. This atomic scattering process does not involve atomic spin interactions, thereby obviating the need to address environmental decoherence effects and eliminating the requirement for a magnetic field to establish a quantization axis in the experiment. The photons scattered by the atom enter the cavity field and are directly coupled into the optical fiber guided mode via one of the microcavity mirrors with a high transmission. The guided out photons contain both elastic and inelastic scattering components. We homemade a narrowband optical notch filter based on the over coupling effect, with a linewidth of $\SI{15}{\mega\hertz}$, allowing us to filter out $\sim 99.5\%$ of the inelastically scattered photons effectively \cite{gallego2016high}. A Hanbury Brown and Twiss (HBT) setup is utilized to measure the second-order correlation function of the photons. By setting the notch filter to be far off-resonant with the driving light, the antibunching of photons $g^{(2)}(0)\approx0.1$ is observed (Fig.\,\ref{fig:HBT}(a)), verifying the single-photon features of the full-spectrum field emitted by a single atom. The Rabi oscillation of the driven atom induces an oscillating pattern with a frequency of $\Omega^\prime=\sqrt{\Omega^2+\Delta_\mathrm{a}^2}$ in the time-resolved correlation function which is dampened over the Purcell-enhanced lifetime $1/\Gamma$. When rejecting the elastic scattered photons by setting the notch filter to be resonant with the driving frequency, a strong photon-bunching $g^{(2)}(0)\approx100$ induced by inelastically scattered photons is observed (Fig.\,\ref{fig:HBT}(c)). By continuously tuning the filtering ratio of the elastic scattered photons, the transformation of optical field from sub-Poissonian to super-Poissonian distribution is demonstrated (Fig.\,\ref{fig:HBT}(a-c)).

%%%%%%%%%%%%%%%%%%%%%%%%%%%%
\begin{figure}[tb]
    \centering
	\includegraphics[width=0.48\textwidth]{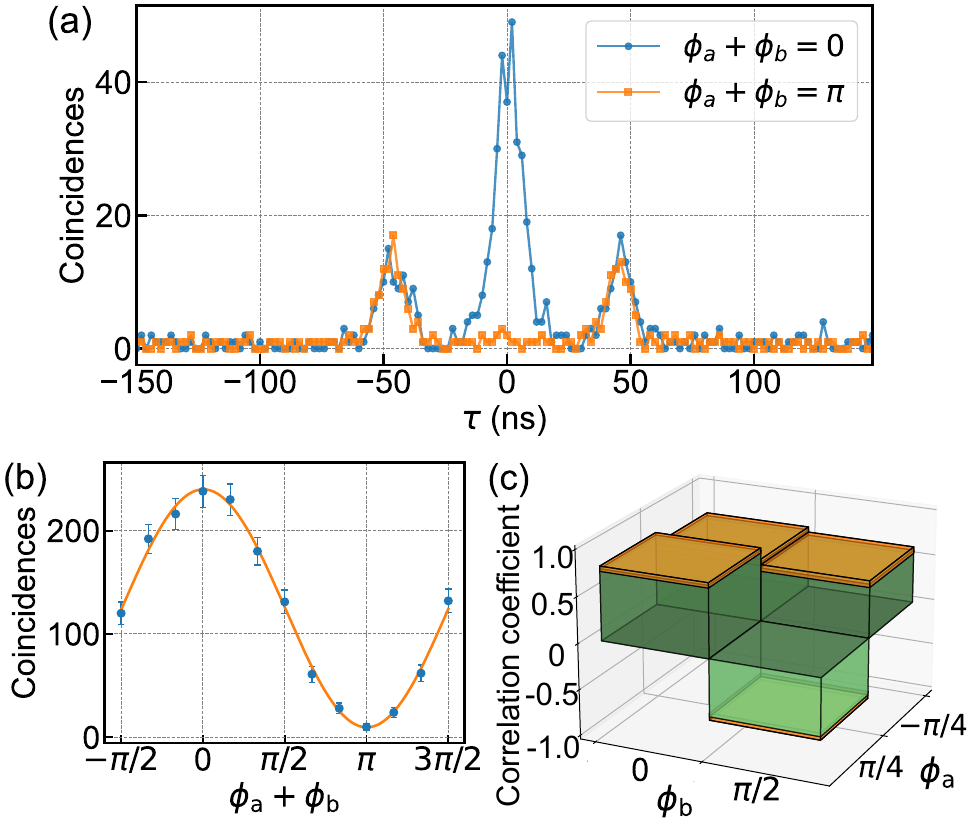}
\caption{Measurements of the Franson interferometer and a violation of the CHSH inequality. (a) Time-resolved correlation measurement between photons passing through the interferometers. (b) The interference curve of coincidences, obtained by scanning $\phi_\mathrm{a}+\phi_\mathrm{b}$. The data integration time are \SI{600}{\second}. The error bars are estimated on the basis of Poisson distribution. (c) A CHSH parameter of $\SI{2.61\pm0.07}{}$ is measured. Green block:correlation coefficient. Yellow block: measurement uncertainty.
}
\label{fig:CHSH}
\end{figure}
%%%%%%%%%%%%%%%%%%%%%%%%%%%%%

To verify the genuine non-local correlations of the two-photon inelastically scattered by single atoms, as depicted in Fig.\,\ref{fig:setup}(d), an Franson-type interferometer is utilized \cite{franson1989bell, jiang2023quantum, ge2024quantum}. Due to the narrow linewidth of atomic energy level, a large delay in the Mach–Zehnder interferometer is required. Therefore, a 9-meter temperature-controlled single-mode fiber is incorporated in the long arm, introducing a delay of \SI{47}{\nano\second}, which is significantly longer than the Purcell-shortened atomic lifetime ($1/\Gamma$). A periodically pulsed, frequency-stabilized \SI{801}{\nano\metre} laser is employed to stabilize the phase differences of the interferometers at $\phi_\mathrm{a}$ ($\phi_\mathrm{b}$). After passing through such Franson interferometer, the energy-time entangled photon pairs are transformed into a Bell state
\begin{equation}\label{equ:path_entangled}
    \frac{1}{\sqrt{2}}\left(\left|s_\mathrm{a},s_\mathrm{b}\right\rangle+\mathrm{e}^{i\left(\phi_\mathrm{a}+\phi_\mathrm{b}\right)}\left|l_\mathrm{a},l_\mathrm{b}\right\rangle\right),
\end{equation}
where $\left|s\right\rangle$ ($\left|l\right\rangle$) corresponds to the state of a photon passing through the short (long) arm. Fig.\,\ref{fig:CHSH}(a) presents two representative measurement results, including two side peaks corresponding to $\left|l_\mathrm{a},s_\mathrm{b}\right\rangle$ and $\left|s_\mathrm{a},l_\mathrm{b}\right\rangle$, and a central peak corresponding to Eq.\,\ref{equ:path_entangled} that exhibits constructive (destructive) interference when $\phi_\mathrm{a}+\phi_\mathrm{b}=0$ ($\pi$). By scanning $\phi_\mathrm{a}+\phi_\mathrm{b}$, the interference pattern shown in Fig.\,\ref{fig:CHSH}(b) is obtained, with each data point obtained by summing the coincidence rates of the central peak within the full width at half maximum of \SI{12}{\nano\second}. A sinusoidal fit indicates a raw interference visibility of $92.6\pm2.3\%$, which is significantly above the $1/\sqrt{2}$ threshold for general entanglement recognition. A definitive proof of entanglement is verified by violating the CHSH Bell inequality. The CHSH parameter $S_\mathrm{CHSH}=\lvert E\left(\phi_\mathrm{a}, \phi_\mathrm{b}\right)-E\left(\phi_\mathrm{a}, \phi_\mathrm{b^\prime}\right)+E\left(\phi_\mathrm{a^\prime}, \phi_\mathrm{b}\right)+E\left(\phi_\mathrm{a^\prime}, \phi_\mathrm{b^\prime}\right)\rvert$ is determined from four correlation coefficients, with each measured on one set of basis $\left(\phi_\mathrm{a}, \phi_\mathrm{b}\right)$ \cite{liu2024violation}. In Fig.\,\ref{fig:CHSH}(c), by setting $\left(\phi_\mathrm{a}, \phi_\mathrm{a^\prime}, \phi_\mathrm{b}, \phi_\mathrm{b^\prime}\right)=\left(\pi/4, -\pi/4, 0, \pi/2\right)$, a definitive violation of the CHSH Bell inequality is observed with $S_\mathrm{CHSH}=\SI{2.61\pm0.07}{} > 2$ of over 8 standard deviations, thereby verifying the entanglement of two-photon inelastically scattered off a single atom with the explicit exclusion of the local hidden-variable.

%%%%%%%%%%%%%%%%%%%%%%%%%%%%
\begin{figure}[tb]
    \centering
        \includegraphics[width=0.46\textwidth]{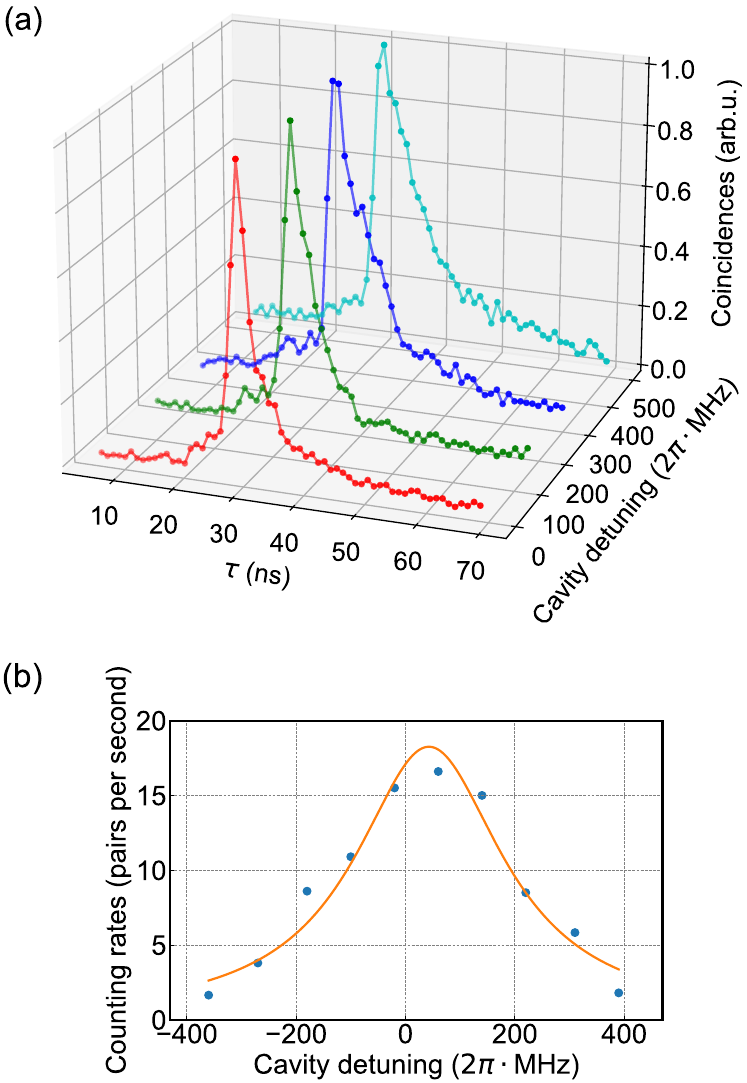}
\caption{(a) Wavepackets of the late photons in the time domain. With increasing cavity detuning, the wavepackets exhibit broadening as a result of the diminished Purcell enhancement. (b) Dependence of the photon pair counting rates on the cavity detuning, with a solid line fitted to Eq.\,\ref{equ:det}.
}
\label{fig:wavepacket} 
\end{figure}
%%%%%%%%%%%%%%%%%%%%%%%%%%%%%

Since the HBT setup disregards the frequency difference of photons, we further employ another homemade optical filter cavity with a linewidth of \SI{150}{\mega\hertz} to separate the photons from distinct sidebands. Fig.\,\ref{fig:wavepacket}(a) presents frequency-separated time-resolved correlation measurements with the detected arrival time of early (late) photons as the \textit{start} (\textit{stop}). The pronounced asymmetry indicates a stringent temporal order between the inelastically scattered photons from two sidebands under large detuning and weak excitation \cite{aspect1980time}. Since early photons act as the triggering signals for latter photons, the distribution of arrival time differences effectively captures the temporal profile of the late photon wavepackets. As the cavity approaches resonant with the atom, the decay time of the photon emission is measured to be \SI{3.7\pm0.3}{\nano\second}, representing a significant compression compared with the theoretically predicted \SI{27.7}{\nano\second} for emission into the free space \cite{schrama1992intensity}. This is in good agreement with the cooperativity of our system ($C=\SI{4.1\pm0.8}{}$), demonstrating that the atom-cavity coupling in the Purcell region also results in a $2C$ enhancement of the inelastically scattered photons. When the cavity is detuned from resonance, the wavepacket compression diminishes due to the reduction in cooperativity, enabling the generation of entangled photon pairs with continuously tunable linewidths \cite{tiecke2014nanophotonic}. This tunability enhances the compatibility of the neutral atom quantum network.

In Fig.\,\ref{fig:wavepacket}(b), we investigate the dependence of the detected photon pair counting rate on the cavity detuning. By fitting the rate data to our model
\begin{equation}\label{equ:det}
	R_\mathrm{det}^{(2)}=\frac{\eta^2}{4}sR_\mathrm{c},
\end{equation} 
we derive an overall photon detection efficiency $\eta$ of $\sim0.03$, which incorporates the FFPC out-coupling efficiency of $\sim0.4$, the optical path transmission efficiency of $\sim0.12$ and the single-photon detector efficiency of $\sim0.6$. When the FFPC is nearly resonant with the driving field, we detect more than 16 photon pairs per second. This corresponds to a total scattering rate of $\sim\SI{36}{\kilo\hertz}$ for entangled photon pairs and a rate of $\sim\SI{5.9}{\kilo\hertz}$ of photon pairs collected into the FFPC out-coupling fiber. With practical improvements such as enhanced fiber coupling, frequency filtering, and precise polarization control to minimize the losses induced by polarization projection, achieving a collected usable photon pair rate exceeding \SI{1}{\kilo\hertz } is attainable for our current system. We have demonstrated remarkably bright photonic Bell state generation with tunable narrowband characteristics.

In conclusion, following the approach of the work \cite{liu2024violation}, we have efficiently produced photonic Bell states through the inelastic scattering of single neutral atoms in the Heiter regime, on the basis of a Purcell-coupling cavity QED platform. The small mode volume of the FFPC results in a strong coupling strength, which permits an enhancement of two-photon scattering process involving two frequency, while maintaining a high cooperativity. We observe a pronounced bunching of inelastically scattered photons and verify the violation of the Bell inequalities through the interference of the photon probability amplitudes, thereby confirming the non-local entanglement properties. Further results demonstrate that the wavepackets of entangled photons can be continuously tuned. Besides, this approach does not involve manipulating of the atomic spin, thereby mitigating the adverse effects of decoherence processes. Our work enhances the understanding of resonance fluorescence in neutral atoms and holds significant potential for advancing future quantum networks.

\begin{acknowledgments}
This work was supported by the Innovation Program for Quantum Science and Technology (No.\,2021ZD0301200), the National Natural Science Foundation of China (No.\,11804330 and No.\,11821404), and the Fundamental Research Funds for the Central Universities (WK2470000038).
\end{acknowledgments}

\bibliography{manuscript}

\clearpage

\part*{\Large\centering Supplementary Information}
\section*{FIBER FABRY-PÉROT CAVITY}
The fiber Fabry-Perot cavity (FFPC) in the experimental apparatus is comprised of two microcavity mirrors, each of which is constructed by a $\mathrm{CO}_2$ laser to machine a concave profile on the fiber end facet, followed by ion beam sputtering coating \cite{pan2022fabrication}. The out-coupling (OC) mirror is fabricated on a single-mode fiber (IVG, Cu800) and features a relatively high transmittance to optimize the collection of photons. On the other side, the high-reflective (HR) mirror is fabricated on a multi-mode fiber (IVG, Cu100/110). The FFPC, with a length of \SI{89.0\pm0.5}{\micro\metre}, features an asymmetric configuration with the radius of curvature of  the OC mirror $R_\mathrm{OC}=\SI{470\pm10}{\micro\metre}$, and the HR mirror $R_\mathrm{HR}=\SI{115\pm5}{\micro\metre}$, to optimize the mode matching from the cavity mode to the guided mode of OC fiber. The cavity field is tightly focused to a radius of \SI{4.53\pm0.04}{\micro\metre} at the center, resulting in greatly enhanced atom-field interactions. The cavity finesse of \SI{5150\pm20}{} leads a field decay rate of $\kappa/2\pi=\SI{164\pm5}{\mega\hertz}$, pushing the coupling between atoms and the cavity into the Purcell regime.

\section*{ATOM LOADING}

To load a single atom into the cavity, we first prepare a small cloud of $^{87}\mathrm{Rb}$ atoms using a magnetic-optical-trap (MOT) with a beam diameter of \SI{2}{\milli\metre}, positioned \SI{3}{\milli\metre} above the FFPC. After \SI{1.5}{\second} of MOT, the magnetic field is turned off and the cooling light detuning increased to apply polarization-gradient-cooling (PGC), cooling the atoms to a temperature of $\sim \SI{20}{\micro\kelvin}$ in \SI{1}{\milli \second}. After release, a \SI{796}{\nano\metre} optical dipole trap (ODT), focused by a in-vacuum aspheric lens ($\mathrm{NA}=0.3$, working distance of \SI{10}{\milli\metre}), guides the atoms along the gravitational ($\hat{z}$) direction toward the center of the FFPC. As the atoms approach the cavity, the \SI{796}{\nano\metre} guiding ODT is switched off and an \SI{850}{\nano\metre} optical lattice is formed, coinciding with the \SI{796}{\nano\metre} ODT but with a smaller waist. Simultaneously, a lin$\perp$lin driving light, angled slightly relative to the  $\hat{z}$ direction and red-detuned by \SI{80}{\mega\hertz} from the unperturbed $F=2\leftrightarrow F'=3$ transition of the $^{87}\mathrm{Rb}$ $\mathrm{D}_2$ line, induces Sisyphus cooling on atoms in all directions, with a primary effect along the $\hat{z}$ direction. Once the kinetic energy of the atoms is removed, they are trapped in a 2D optical lattice ($U_0/k_\mathrm{B}\sim\SI{0.6}{\milli\kelvin}$) formed by the \SI{850}{\nano\metre} optical lattice mentioned above and an \SI{801}{\nano\metre} intracavity optical lattice \cite{pan2022feedback} along the $\hat{x}$ direction.

%%%%%%%%%%%%%%%%%%%%%%%%%%%%
\begin{figure}
	\includegraphics[width=0.46\textwidth]{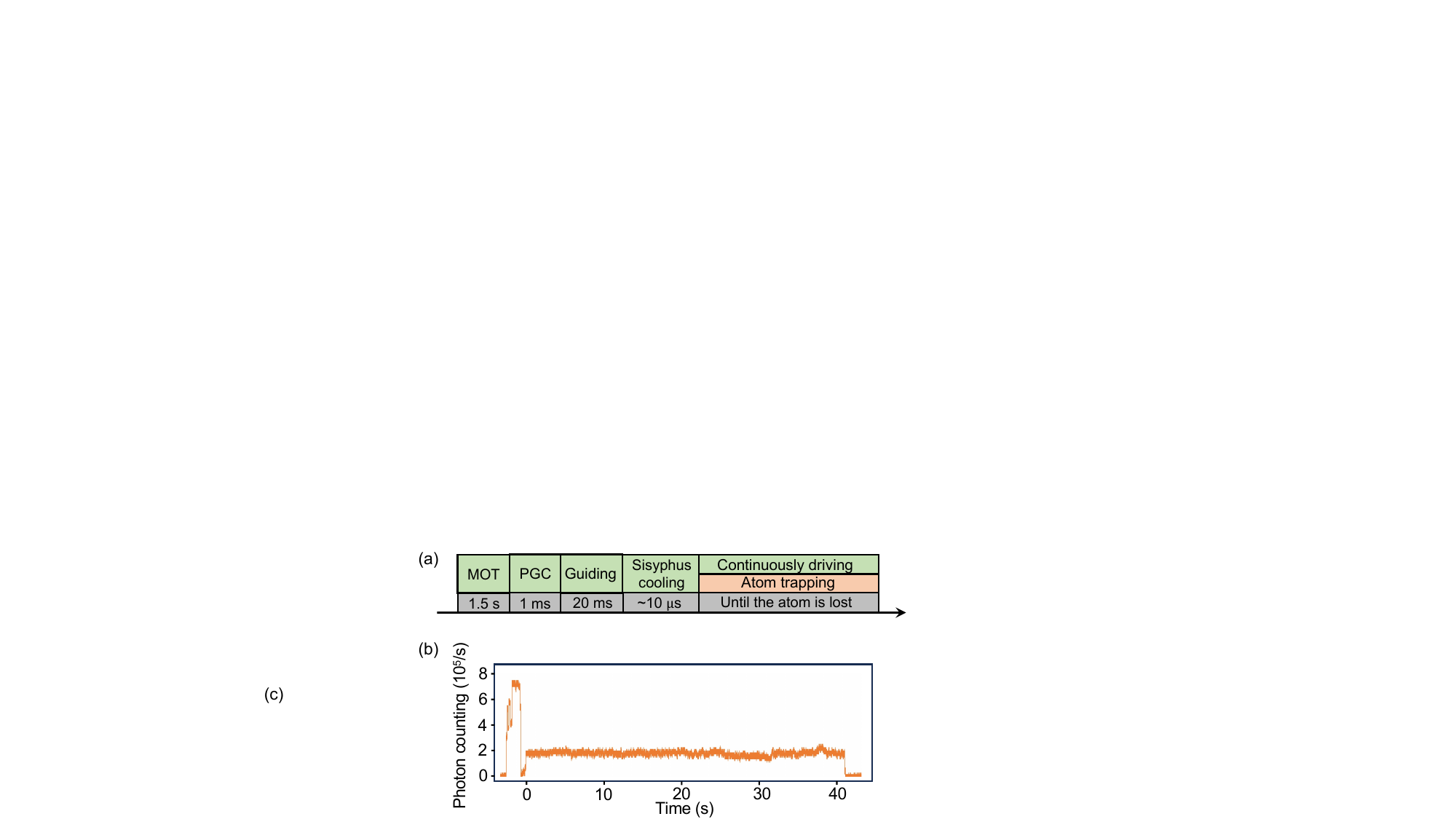}
	\caption{\label{fig:supp_atom loading} (a) Sketch of experimental apparatus, not to scaled. (b) Experimental sequence for atom loading and driving. (c) Photon counting detected through the FFPC. Integral time: \SI{10}{\milli\second}.
	}
\end{figure}
%%%%%%%%%%%%%%%%%%%%%%%%%%%%%

Photon counting of the driving light scattering into the cavity serves as a tool for verifying single-atom confinement. Fig.\,\ref{fig:supp_atom loading}(b) is a typical photon counting trace detected through the FFPC. For $t<0$, multiple atoms are trapped simultaneously, leading to shorter lifetimes. For $t>0$, a single atom is trapped, which has a longer lifetime ($\tau_{1/\mathrm{e}}\approx\SI{20}{\second}$) and triggers the commencement of the subsequent experiments. In the experiment, each release of an atomic cloud from the MOT results in an 80\% probability of single-atom trapping.

\section*{FREQUENCY-FILTERED PHOTON SECOND-ORDER CORRELATIONS }

%%%%%%%%%%%%%%%%%%%%%%%%%%%%
\begin{figure}
	\includegraphics[width=0.483\textwidth]{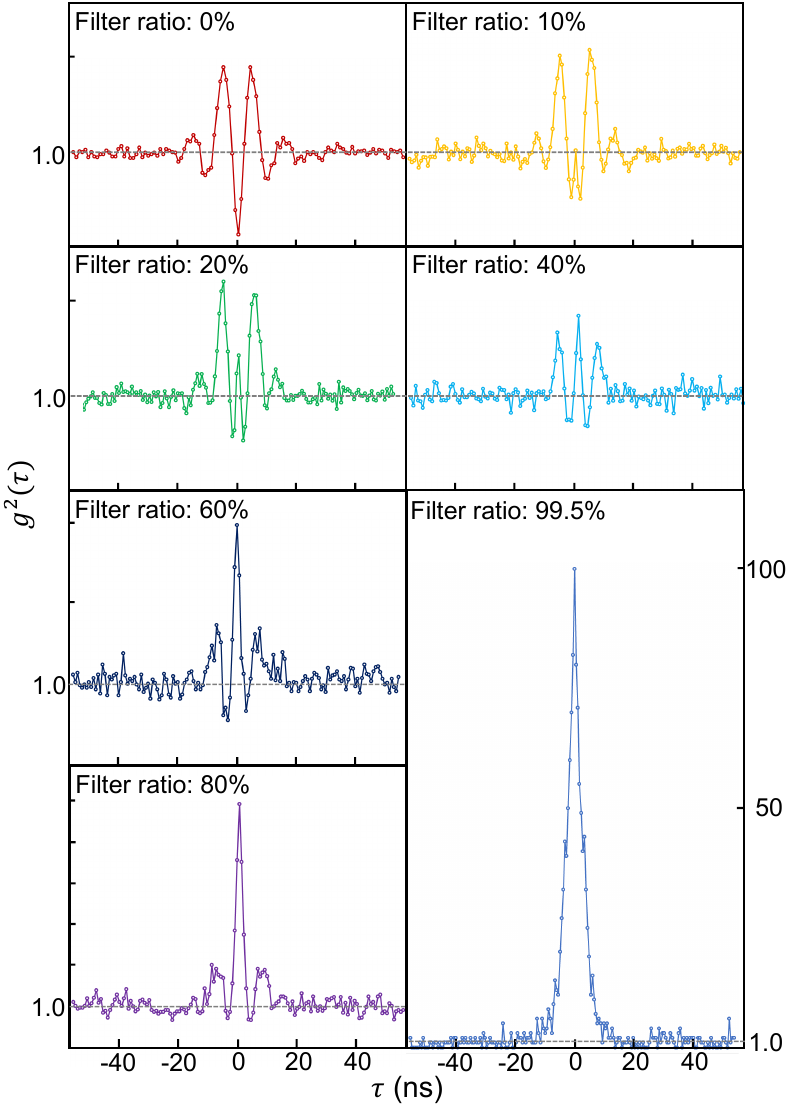}
	\caption{\label{fig:supp_HBT} Second-order correlation functions with different filtering ratios of elastically scatterred photons.
	}
\end{figure}
%%%%%%%%%%%%%%%%%%%%%%%%%%%%%

To filter out elastically scattered photons at the same frequency with the driving light, a narrow-band notch filter composed of an optical cavity with a length of \SI{50}{\milli\metre} and a finesse of 200 is used. The cavity length is actively stabilized using the Pound-Drever-Hall method to a laser that is frequency-locked to an optical frequency comb (FC1500-250-ULN, Menlo Systems) \cite{shen2023continuously}. When the incident light frequency matches the filtering cavity’s resonance, the reflected field comprises two components: the directly reflected field and the leakage field from the cavity circulating field. The destructive interference between these two components results in a reduction in the reflected light intensity. In our apparatus, the transmittance of the incident mirror is greater than that of the transmissive mirror. By precisely adjusting the angle of the transmissive mirror and coupling the reflected light into a single-mode fiber, the coupled amplitudes of the two field components can be adjusted nearly equal, achieving near-complete destructive interference \cite{gallego2016high}. This allows for the filtering of $\sim99.5\%$ of the light at the resonance point.

In the experiments, the filtering cavity is locked to the laser sideband generated by an electro-optic modulator. By tuning the resonant frequency of the filter cavity, different ratios of inelastically scattered photons can be selectively filtered. In Fig.\,\ref{fig:supp_HBT}, we present the second-order correlation function of the field output from the FFPC under varying filtering ratios. The results indicate that as the filtering ratio increased from 0, the light field transformed from a sub-Poissonian distribution featuring anti-bunching to a super-Poissonian distribution featuring pronounced bunching.

\section*{FRANSON INTERFEROMETER}
%%%%%%%%%%%%%%%%%%%%%%%%%%%%%%%%%%%%%%%%%%%%%%%%%%%%%%%%%%%%%%%%%%%
\begin{figure*}[tb]
    \centering
        \includegraphics[width=1.0\textwidth]{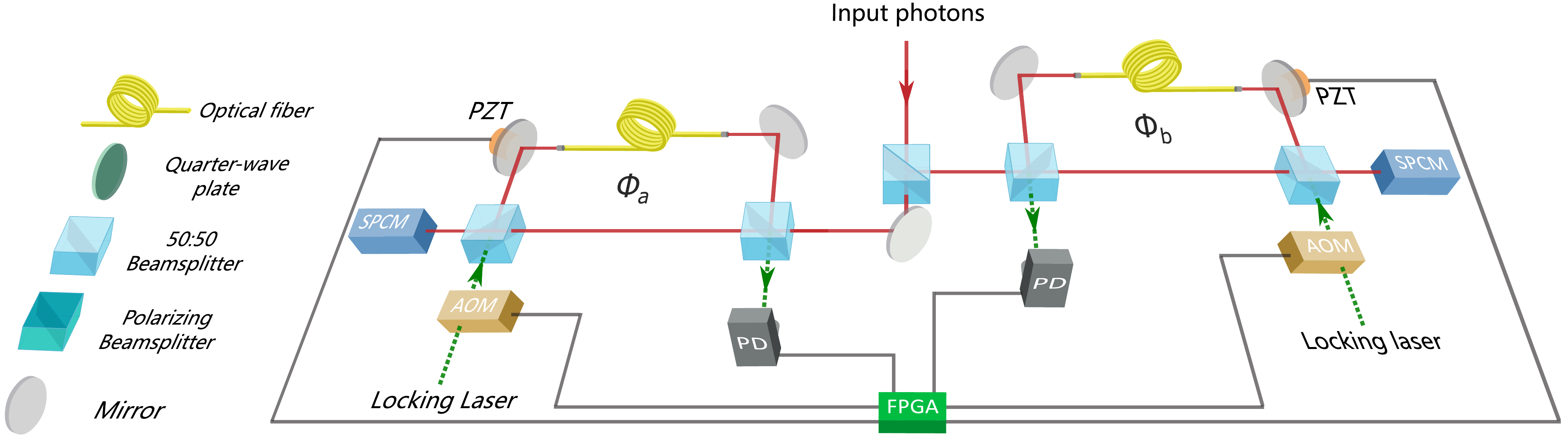}
\caption{
   Franson interferometer for photon energy-time entanglement verification. Two identical unbalanced Mach–Zehnder (M-Z) interferometers each with a \SI{9}{\metre} temperature-controlled optical fiber in the long arm. An Acousto-optic modulators (AOMs) periodically activate the locking laser to stabilize the phase difference of the interferometer, with the process managed by a field-programmable gate array (FPGA). PD, photodiode; SPCM, single-photon counting module.}
\label{fig:supp_setup} 
\end{figure*}

%%%%%%%%%%%%%%%%%%%%%%%%%%%%%%%%%%%%%%%%%%%%%%%%%%%%%%%%%%%%%%%%%%%
Compared to quantum dots, neutral atoms exhibit an exceptionally narrow linewidth (in the $\sim\SI{}{\mega\hertz}$ range). Consequently, Franson interferometer measurements in atomic systems require a longer optical path difference between the long and short arms, which is challenging to achieve in free space. In this experiment,as depicted in Fig.\,\ref{fig:supp_setup}, we utilized a single-mode fiber-based interferometer to introduce the optical path difference, with a 9-meter fiber introducing a phase difference of 47 ns. The optical fibers are actively temperature-stabilized using thermoelectric coolers (TECs) and isolated from vibrations. The phase difference of each UMZI is actively stabilized and locked to the reference \SI{801}{\nano\metre} light using a mirror equipped with a piezoelectric transducer (PZT). To mitigate the impact of photons generated by Raman scattering in the fiber, which can severely degrade the signal-to-noise ratio, the reference light is modulated into periodic pulses in the time domain by AOMs.

%%%%%%%%%%%%%%%%%%%%%%%%%%%%
\begin{figure}
	\includegraphics[width=0.48\textwidth]{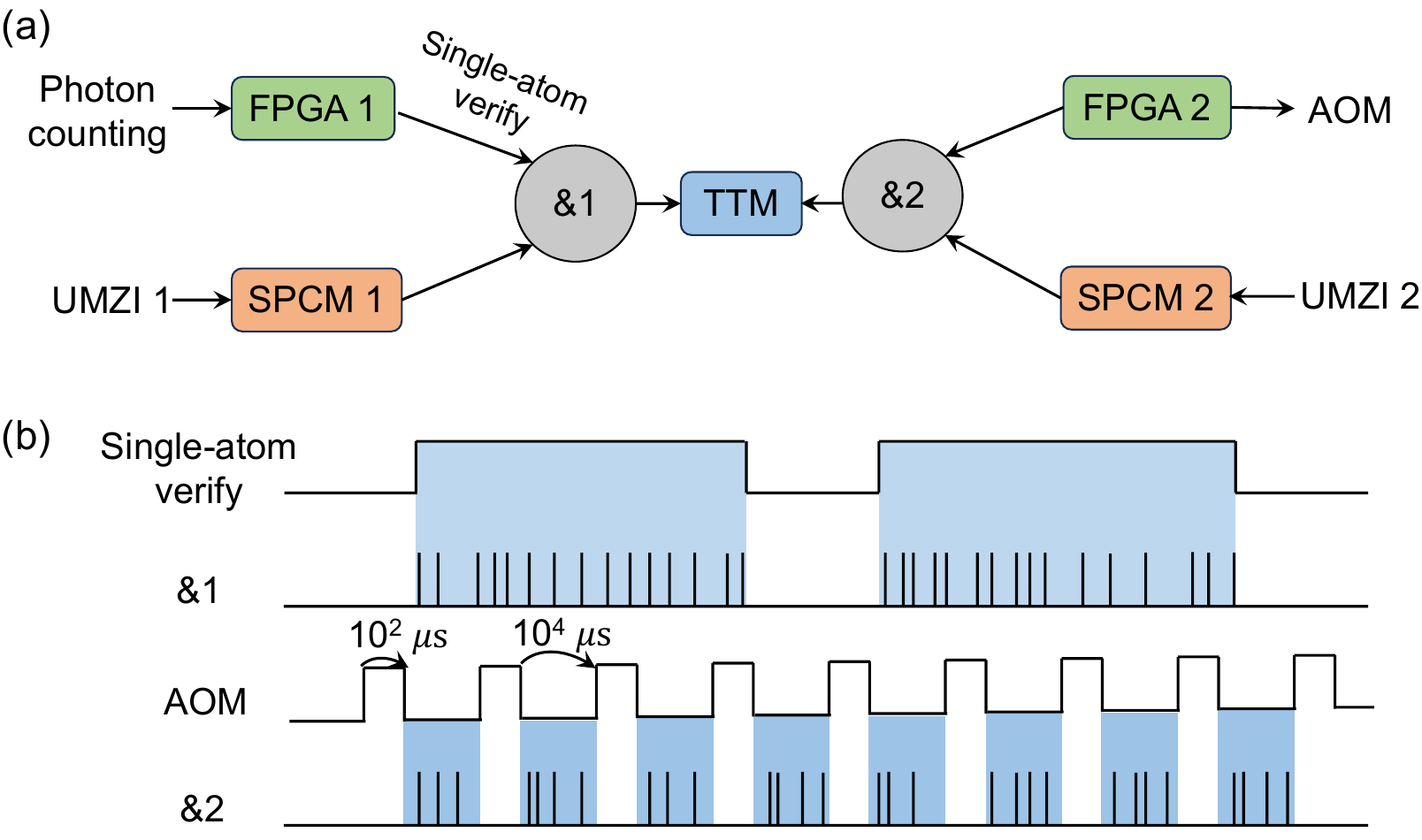}
	\caption{\label{fig:supp_Franson} (a) Schematic diagram of signals used in Franson interferometer. (b) Time sequence of Franson interferomter measurement. FPGA, field-programmable gate array; UMZI, unbalanced Mach-Zehnder inteferometer.; \&, high-speed AND gate; TTM, time-tagger module; SPCM, single-photon counting module; AOM, acoustic-optic modulator.
	}
\end{figure}
%%%%%%%%%%%%%%%%%%%%%%%%%%%%%

As depicted in Fig.\,\ref{fig:supp_Franson}(a), two FPGAs were employed to manage logic operations and time sequences. FPGA1 processes photon sampling signals to generate single-atom indicating signals, which are fed into a high-speed AND gate along with photon signals recorded by SPCM1. FPGA2 outputs the control signal for the AOM, which, after inversion, is combined with the photon signal from SPCM2 in another high-speed AND gate. The outputs from two AND gates are then time-tagged by a TTM. The time sequences are show in Fig.\,\ref{fig:supp_Franson}(b), AOM controls the input of reference light and opens for $\SI{100}{\micro\second}$ every $\SI{10}{\milli\second}$ for active stabilization of the phase difference.

\section*{CHSH INEQUALITY}

The Clauser-Horn-Shimony-Holt (CHSH) Bell inequality provides a means to demonstrate that certain correlations predicted by quantum entanglement cannot be accounted for by any local hidden-variable theories. The CHSH parameter is determined by the correlation coefficients measured across four distinct measurement basis $S_\mathrm{CHSH}=\lvert E\left(\phi_\mathrm{a}, \phi_\mathrm{b}\right)-E\left(\phi_\mathrm{a}, \phi_\mathrm{b^\prime}\right)+E\left(\phi_\mathrm{a^\prime}, \phi_\mathrm{b}\right)+E\left(\phi_\mathrm{a^\prime}, \phi_\mathrm{b^\prime}\right)\rvert$. Each correlation coefficient is determined by the number of coincidences $n_{ij}\left(\phi_\mathrm{a},\phi_\mathrm{b}\right)$ between Alice's detector $i$ and Bob's detector $j$ for a given pair of measurement basis $\left(\phi_\mathrm{a},\phi_\mathrm{b}\right)$
\begin{multline}\label{equ:supp_1}
    E\left(\phi_\mathrm{a}, \phi_\mathrm{b}\right)=\\
    \frac{n_{11}\left(\phi_\mathrm{a},\phi_\mathrm{b}\right) + n_{22}\left(\phi_\mathrm{a},\phi_\mathrm{b}\right) 
    - n_{12}\left(\phi_\mathrm{a},\phi_\mathrm{b}\right) - n_{21}\left(\phi_\mathrm{a},\phi_\mathrm{b}\right)}{n_{11}\left(\phi_\mathrm{a},\phi_\mathrm{b}\right) + n_{22}\left(\phi_\mathrm{a},\phi_\mathrm{b}\right) 
    + n_{12}\left(\phi_\mathrm{a},\phi_\mathrm{b}\right) + n_{21}\left(\phi_\mathrm{a},\phi_\mathrm{b}\right)}.
\end{multline}
%%%%%%%%%%%%%%%%%%%%%%%%%%%%%%%%%%%%%%%%%%%%
\begin{table}[h!]
    \centering
    \begin{tabular}{|c|c|c|c|c|}
        \hline
        \diagbox{$\phi_\mathrm{b}$}{$\phi_\mathrm{a}$} & $-\pi/4$ & $\pi/4$ & $3\pi/4$ & $5\pi/4$ \\
        \hline
        0& 190 & 205 & 46 & 40 \\
        \hline
        $\pi/2$ & 256 & 38 & 24 & 160\\
        \hline
        $\pi$ & 60 & 56 &206 & 239 \\
        \hline
        $3\pi/2$ & 39 & 209 & 220 & 49\\
        \hline
    \end{tabular}
    \caption{CHSH parameter measurements.}
    \label{tab:CHSH}
\end{table}
%%%%%%%%%%%%%%%%%%%%%%%%%%%%%%%%%%%%%%%%%%%%%
In the case where Alice and Bob each have only one detector, Eq.\,\ref{equ:supp_1} can be equivalently implemented by expanding the number of measurement basis
\begin{multline}\label{equ:supp_2}
    E\left(\phi_\mathrm{a}, \phi_\mathrm{b}\right)=\\
    \frac{n\left(\phi_\mathrm{a},\phi_\mathrm{b}\right) + n\left(\phi_\mathrm{a\perp},\phi_\mathrm{b\perp}\right) 
    - n\left(\phi_\mathrm{a},\phi_\mathrm{b\perp}\right) - n\left(\phi_\mathrm{a\perp},\phi_\mathrm{b}\right)}{n\left(\phi_\mathrm{a},\phi_\mathrm{b}\right) + n\left(\phi_\mathrm{a\perp},\phi_\mathrm{b\perp}\right) 
    + n\left(\phi_\mathrm{a},\phi_\mathrm{b\perp}\right) + n\left(\phi_\mathrm{a\perp},\phi_\mathrm{b}\right)},
\end{multline}
where $\phi_\mathrm{a\perp}$ ($\phi_\mathrm{b\perp}$) is the orthogonal basis of
$\phi_\mathrm{a}$ ($\phi_\mathrm{b}$).

In our measurement, the basis $\left(\phi_\mathrm{a}, \phi_\mathrm{a^\prime}, \phi_\mathrm{b}, \phi_\mathrm{b^\prime}\right)$ were configured as $\left(\pi/4, -\pi/4, 0, \pi/2\right)$ and $\phi_\mathrm{a\perp}=\phi_\mathrm{a}+\pi$, $\phi_\mathrm{b\perp}=\phi_\mathrm{b}+\pi$. The coincidence results are presented in Tab.\,\ref{tab:CHSH}, leading to a conclusion of $S_\mathrm{CHSH}=\SI{2.61\pm0.07}{}$, with the uncertainty derived from the Poisson distribution.

\end{document}